\documentstyle[12pt,subeq]{article}

\title{Evolution of quantum systems with a scaling type of 
time-dependent Hamiltonians}

\author{L. \v SAMAJ \\ \\
Institute of Physics, Slovak Academy of Sciences, \\
D\'ubravsk\'a cesta 9, 842 28 Bratislava, Slovakia; \\
tel. +421 2 5941 0522, fax +421 2 5477 6085 \\
e-mail: fyzimaes@savba.sk}

\begin{document}

\maketitle

\begin{abstract}
We introduce a new class of quantum models with
time-dependent Hamiltonians of a special scaling form.
By using a couple of time-dependent unitary transformations, 
the time evolution of these models is expressed in terms of 
related systems with time-independent Hamiltonians.
The mapping of dynamics can be performed in any dimension, for an
arbitrary number of interacting particles and for any type of
the scaling interaction potential.
The exact solvability of a ``dual'' time-independent Hamiltonian 
automatically means the exact solvability of the original problem 
with model time-dependence.
\end{abstract}

\vfill



\newpage

Dynamics of quantum systems, governed by time-dependent Hamiltonians,
has attracted much of attention for a long time.
A relevant progress has been made mainly in the study of one-dimensional
time-dependent harmonic oscillators$^{1-10}$
which have many applications in various areas of physics
(see e.g. Refs. 11-13).
Lewis$^1$ and Lewis and Riesenfeld$^2$ (LR)
have introduced for these systems a quantum-mechanical LR invariant,
and derived a relation between the invariant eigenstates and exact
solutions of the corresponding time-dependent Schr\"odinger equation.
Using the LR invariant method, exact wavefunctions have been obtained
for harmonic oscillators with time-dependent frequency,$^{3-5}$ 
time-dependent mass and frequency,$^{6,7}$ and linear driving terms.$^8$
At present, the exact solution can be, in principle, constructed 
for a general one-dimensional time-dependent Hamiltonian of $N$ 
coupled quantum oscillators.$^{9,10}$
The addition of a singular inverse quadratic potential does not
break the exact solvability of the time-dependent harmonic oscillator,
as has been shown by combining the LR invariant method with a unitary
transformation in Refs. 14-17.

In this paper, we present a new vast array of quantum models with
time-dependent Hamiltonians of a special scaling form.
By using a couple of time-dependent unitary transformations, 
time evolution of these models is expressed in terms of 
related systems with time-independent Hamiltonians.
The mapping of dynamics can be performed in any dimension, for an
arbitrary number of interacting particles and for any type of the
interaction potential.
The exact solvability of the time-independent Hamiltonian automatically
means the exact solvability of the original problem with model 
time-dependence; otherwise, dynamics induced by the time-dependent 
Hamiltonian can be studied by simpler techniques known for 
time-independent Hamiltonians.

We first restrict ourselves to the case of one particle situated
in a $d$-dimensional space of points ${\bf r}=(x_1,x_2,\ldots,x_{d})$.
The proposed time-dependent Hamiltonian has the following scaling form:
\begin{equation} \label{1}
{\bf \hat H}(t) = {f'(t) \over f'(0)} \left[
{f(t)\over f(0)} {{\bf \hat p}^2 \over 2 m} 
+ {f(0)\over f(t)}~ V\left( {f(0)\over f(t)}{\bf r}\right) \right] ,
\end{equation}
where ${\bf \hat p} = - {\rm i}\hbar \nabla_{{\bf r}}$ is the momentum
operator, $V$ is an arbitrary one-body potential and the prime
denotes the differentiation with respect to the argument.
The normalization of parameters was chosen such that at initial 
time $t=0$ the Hamiltonian takes the standard form
\begin{equation} \label{2}
{\bf \hat H}(t=0) = {{\bf \hat p}^2 \over 2 m} + V({\bf r}) .
\end{equation}
The function $f(t)$ is assumed to be real continuous function of time 
in order to ensure the hermiticity of the Hamiltonian.
To simplify the notation, without any lost of generality we shall
assume that
\begin{equation} \label{3}
f(0) = 1 , \quad f'(0) = \epsilon \ (\epsilon \ {\rm real\ and\ nonzero})
\end{equation} 
and rewrite the Hamiltonian (\ref{1}) as follows
$${\bf \hat H}(t) = {f'(t) \over \epsilon} \left[
f(t) {{\bf \hat p}^2 \over 2 m} + {1\over f(t)}~ 
V\left( {{\bf r}\over f(t)} \right) \right] . \eqno(1') 
$$
Two choices of the function $f(t)$ are of special interest.
The first one,
\begin{equation} \label{4}
f(t)  =  (1 + 2 \epsilon t)^{1/2}
\end{equation}
with $\epsilon > 0$ in order to avoid singularities in time,
corresponds to the constant particle mass.
The corresponding Hamiltonian (\ref{1}) takes the form
\begin{equation} \label{5}
{\bf \hat H}(t) = {{\bf \hat p}^2\over 2 m} + {1\over 1 + 2 \epsilon t} 
V\left({{\bf r}\over (1+2\epsilon t)^{1/2}}\right) .
\end{equation}
The classical version of this Hamiltonian has been introduced in Ref. 18
in connection with the study of adiabatic propagation
of distributions, with $\epsilon$ taken as the adiabatic slowness parameter.
The second choice is
\begin{equation} \label{6}
f(t) =   \exp ( \epsilon t ) ,
\end{equation}
and the corresponding Hamiltonian reads
\begin{equation} \label{7}
{\bf \hat H}(t) = {\rm e}^{2 \epsilon t} {{\bf \hat p}^2\over 2 m} 
+ V\left({\rm e}^{-\epsilon t}{\bf r}\right) .
\end{equation}
In one dimension and when $V(x) = m \omega^2 x^2 / 2$, the model (\ref{7})
is known as the Caldirola-Kanai oscillator.$^{19-21}$
Its exact quantum states are known, even in a more general case
with the singular inverse square potential$^{16}$,
$V(x) = m \omega^2 x^2 / 2 + k / x^2$.

The evolution of the system with Hamiltonian $(1')$
is governed by the time-dependent Schr\"odinger equation
\begin{equation} \label{8}
{\rm i} \hbar {\partial \over \partial t} \psi({\bf r},t)
= {\bf \hat H}(t) \psi({\bf r},t) ,
\end{equation}
with the initial condition for the wavefunction
\begin{equation} \label{9}
\psi({\bf r},t=0) = \psi({\bf r}) .
\end{equation}
In the next two paragraphs, we apply successively two time-dependent 
unitary transformations, 
$$
\{ {\bf \hat H}(t), \psi({\bf r},t) \} \Longrightarrow 
\{ {\bf \hat H}_1, \psi_1({\bf r},t) \} \Longrightarrow 
\{ {\bf \hat H}_2, \psi_2({\bf r},t) \} , 
$$
to go from the original time-dependent Hamiltonian ${\bf \hat H}(t)$
to the new time-independent Hamiltonian ${\bf \hat H}_2$, via 
an intermediate one ${\bf \hat H}_1$.

We start with the unitary transformation
\begin{subequations} 
\begin{eqnarray}
t' & = & {1\over \epsilon} \ln f(t) , \label{10a} \\
x'_j & = & {x_j \over f(t)} ,
\quad \quad j = 1, 2, \ldots, d . \label{10b}
\end{eqnarray}
\end{subequations}
It is easy to verify that it holds
\begin{subequations} 
\begin{eqnarray}
{\partial \over \partial t} & = & 
{f'(t)\over f(t)} \left[ {1\over \epsilon} {\partial \over \partial t'}
- {\bf r}'\cdot \nabla_{{\bf r}'} \right] ,
\label{11a}\\ 
{\partial^2 \over \partial x_j^2} & = & 
{1 \over [f(t)]^2} {\partial^2 \over \partial {x'_j}^2} ,
\quad \quad j = 1, 2, \ldots, d . \label{11b}
\end{eqnarray}
\end{subequations}
The transformation (10) converts the evolution equation
(\ref{8}) to the new one
\begin{subequations} 
\begin{eqnarray}
{\rm i} \hbar {\partial \over \partial t} \psi_1({\bf r},t)
& = & {\bf \hat H}_1 \psi_1({\bf r},t) , \label{12a} \\
{\bf \hat H}_1 & = & {{\bf \hat p}^2\over 2 m} - 
\epsilon {\bf r}\cdot {\bf \hat p} + V({\bf r}) .\label{12b}
\end{eqnarray}
\end{subequations}
The original wavefunction $\psi$ in (\ref{8}) is expressible in terms
of $\psi_1$ as follows
\begin{equation} \label{13}
\psi({\bf r},t) = \psi_1\left( {{\bf r}\over f(t)},
{1\over \epsilon} \ln f(t) \right) ,
\end{equation}
and the initial condition (\ref{9}) now reads
\begin{equation} \label{14}
\psi_1({\bf r},t=0) = \psi({\bf r}) .
\end{equation}

The time-independent Hamiltonian ${\bf \hat H}_1$ in (12)
can be reexpressed as follows
\begin{eqnarray} \label{15}
{\bf \hat H}_1 & = & 
{1\over 2m} \left( {\bf \hat p} - m \epsilon {\bf r} \right)^2
+ {1\over 2} \epsilon
\left( {\bf \hat p}\cdot {\bf \hat r} - {\bf \hat r}\cdot {\bf \hat p} \right)
\nonumber \\
& & + V({\bf r}) - {1\over 2} m \epsilon^2 r^2 .
\end{eqnarray}
Since
\begin{equation} \label{16}
{\bf \hat p}\cdot {\bf \hat r} = {\bf \hat r}\cdot {\bf \hat p} - 
{\rm i}\hbar d ,
\end{equation}
the evolution Eq. (12) can be written as
\begin{eqnarray} \label{17}
{\rm i}\hbar \left( {\partial\over \partial t} 
+ {d \epsilon \over 2} \right)
\psi_1({\bf r},t) & = & {1\over 2m} 
\left( {\bf \hat p} - m \epsilon {\bf r} \right)^2
\psi_1({\bf r},t) \nonumber \\
& & + \left[ V({\bf r}) - {1\over 2}m \epsilon^2
r^2 \right] \psi_1({\bf r},t) .
\end{eqnarray}
Using the unitary transformation
\begin{equation} \label{18}
\psi_1({\bf r},t) = \exp \left( - {d \epsilon \over 2} t
+ {{\rm i} m \epsilon \over 2\hbar} r^2 \right) \psi_2({\bf r},t)
\end{equation}
we finally arrive at
\begin{subequations} 
\begin{eqnarray}
{\rm i}\hbar {\partial \over \partial t} \psi_2({\bf r},t) 
& = & {\bf \hat H}_2 \psi_2({\bf r},t) , \label{19a} \\
{\bf \hat H}_2 & = & {{\bf \hat p}^2\over 2m} + 
\left[ V({\bf r}) - {1\over 2} m \epsilon^2 r^2 \right] . \label{19b}
\end{eqnarray}
\end{subequations}
The initial condition for $\psi_2$ is implied by Eqs. (\ref{14}) and
(\ref{18}) as follows
\begin{equation} \label{20}
\psi_2({\bf r},t=0) = \psi({\bf r}) \exp\left( - 
{{\rm i} m \epsilon \over 2\hbar} r^2 \right) .
\end{equation}
The original wavefunction $\psi$ is obtained in terms of $\psi_2$
by combining relations (\ref{13}) and (\ref{18}),
\begin{equation} \label{21}
\psi({\bf r},t)  =  {1\over [f(t)]^{d/2}} 
\exp\left( {{\rm i} m \epsilon\over 2\hbar}
\left[ {r\over f(t)}\right]^2 \right) 
\ \psi_2\left( {{\bf r}\over f(t)}, {1\over \epsilon}
\ln f(t) \right) .
\end{equation}

Using the above mapping of dynamics, standard techniques
able to deal with time-independent Hamiltonians (see e.g. Ref. 22)
can be directly applied to time-dependent Hamiltonians.
The mapping becomes even more appealing when the resulting
time-independent problem with potential 
$$ V({\bf r}) - {1\over 2} m \epsilon^2 r^2$$
is exactly solvable, because this automatically means the exact solvability 
of the original model time-dependence.
It is clear that $V({\bf r})$ must be expressible as an integrable
potential plus the harmonic-oscillator potential
$m\epsilon^2 r^2/2$.
In one dimension, there exists an infinite chain of exactly solvable 
``reflectionless potentials''.$^{23,24}$
In higher dimensions, there are standard models which admit
the exact solution.$^{22}$
We briefly discuss the simplest example -- 
the isotropic $d$-dimensional harmonic oscillator with
$V({\bf r}) = m \omega^2 r^2/2$.
For the special choices (\ref{4}) and (\ref{6}) of the function $f(t)$, 
the respective time-dependent Hamiltonians
\begin{equation} \label{22}
{\bf \hat H}(t) = {{\bf \hat p}^2\over 2m} + {1\over 2} m
\left( {\omega\over 1+2\epsilon t} \right)^2 r^2 
\end{equation}
and
\begin{equation} \label{23}
{\bf \hat H}(t) = {\rm e}^{2\epsilon t} {{\bf \hat p}^2\over 2 m}
+ {1\over 2} {\rm e}^{-2\epsilon t} m \omega^2 r^2
\end{equation}
are transformed to the same time-independent Hamiltonian
\begin{equation} \label{24}
{\bf \hat H}_2 = {{\bf \hat p}^2\over 2m} + {1\over 2} m
(\omega^2 - \epsilon^2) r^2 .
\end{equation}
The time evolution associated with this Hamiltonian can be
obtained by using the method of the separation of variables.
Without going into details, there are three regimes: 
if $\omega > \epsilon$ one expands the
wavefunction $\psi_2$ (and, consequently, $\psi$) in Hermite
polynomials, if $\omega = \epsilon$ in plane waves and if
$\omega < \epsilon$ in hypergeometric functions.

The extension of the formalism to the case of an arbitrary
number of particles is straightforward and we only write down
final formulae.
For $N$ particles, the Hamiltonian $(1')$ generalizes to
\begin{equation} \label{25}
{\bf \hat H}(t) = {f'(t)\over \epsilon} \left[
f(t) \sum_{j=1}^N {{\bf \hat p}^2\over 2 m_j} + {1\over f(t)} ~
V\left( \left\{ {{\bf r}_j\over f(t)} \right\} \right) \right] ,
\end{equation}
where $f(0)=1$ and $f'(0)=\epsilon$, and the interaction function
$V$ [with particle coordinates uniformly scaled by $1/f(t)$]
involves all possible one-, two-, $\ldots$, $N$-body potentials.
The solution of the time-dependent Schr\"odinger equation
\begin{equation} \label{26}
{\rm i}\hbar {\partial \over \partial t} \psi(\{ {\bf r}_j\},t) 
= {\bf \hat H}(t) \psi(\{ {\bf r}_j\},t)
\end{equation}
with the initial condition
\begin{equation} \label{27}
\psi(\{ {\bf r}_j\},t=0) = \psi(\{ {\bf r}_j\})
\end{equation}
reads
\begin{equation} \label{28}
\psi(\{ {\bf r}_j\},t) = {1\over [f(t)]^{N d/2}}
\exp \left( {{\rm i}\epsilon \over 2\hbar}
\sum_{j=1}^N m_j \left[ { r_j\over f(t)} \right]^2 \right)
~ \psi_2 \left( \left\{ {{\bf r}_j\over f(t)} \right\},
{1\over \epsilon} \ln f(t) \right) .
\end{equation}
Here, the wavefunction $\psi_2$ evolves under the action
of a time-independent Hamiltonian,
\begin{subequations} 
\begin{eqnarray}
{\rm i}\hbar {\partial \over \partial t} \psi_2(\{ {\bf r}_j\},t) 
& = & {\bf \hat H}_2 \psi_2(\{ {\bf r}_j\},t) , \label{29a} \\
{\bf \hat H}_2 & = & \sum_{j=1}^N {{\bf \hat p}_j^2\over 2m_j} + 
\left[ V({\bf r}) - \sum_{j=1}^N {1\over 2} m_j \epsilon^2 r_j^2 \right] , 
\label{29b}
\end{eqnarray}
\end{subequations}
with the initial condition
\begin{equation} \label{30}
\psi_2(\{ {\bf r}_j\},t=0) = \psi(\{ {\bf r}_j\}) 
\exp\left( - {{\rm i} \epsilon \over 2\hbar} 
\sum_{j=1}^N m_j r_j^2 \right) .
\end{equation}
As the simplest exactly solvable example, we mention a system of $N$ 
coupled $d$-dimensional harmonic oscillators, which is solvable 
in the transformed picture with the time-inde\-pen\-dent Hamiltonian 
by the ordinary normal-mode technique.

The couple of time-dependent unitary transformations presented
in this paper can be applied directly to the quantum Liouville
equation for the density matrix, extending in this way the treatment 
also to the propagation of distributions.

In conclusion, we believe that the results presented will enable one to
answer some problems concerning the evolution of quantum systems under
a time-dependent Hamiltonian, for instance, to predict a 
qualitative change in the analytic structure of leading corrections
to the ideal adiabaticity when going from a few degrees of freedom
to a dissipative system with infinite degrees of freedom.
For classical adiabatic processes in one dimension, 
the corresponding exact analysis has been done in Ref. 17.

\vskip 2truecm

\noindent {\bf Acknowledgements}: 
I am grateful to Prof. J. K. Percus for stimulating discussions.
This work was supported by Grant VEGA 2/7174/20.

\end{document}